# Extremely High Q-factor metamaterials due to Anapole Excitation

Alexey A. Basharin[1], Vitaly Chuguevsky[1,2], Nikita Volsky[1], Maria Kafesaki[3], Eleftherios N. Economou[3]

*1 National University of Science and Technology MISiS, Moscow, 119049, Russia*
*2 Voronezh State Technical University, Voronezh, 394026 Russia*
*3 Foundation for Research and Technology Hellas (FORTH), and University of Crete, Heraklion, Crete, Greece*

## ABSTRACT

We demonstrate that ideal anapole metamaterials have infinite Q-factor. We have designed and fabricated a metamaterial consisting of planar metamolecules which exhibit anapole behavior in the sense that the electric dipole radiation is almost cancelled by the toroidal dipole one, producing thus an extremely high Q-factor at the resonance frequency. The size of the system, at the mm range, and the parasitic magnetic quadrupole radiation are the factors limiting the size of the Q-factor. In spite of the very low radiation losses the local fields at the metamolecules are extremely high, of the order of $10^4$ higher than the external incoming field.

## I. INTRODUCTION

Static toroidal dipole moment, also known as anapole, was firstly introduced by Zel'dovich in 1958 [1]. Its importance is widely recognized in nuclear, molecular and atomic physics [2]. Zel'dovich was able to explain parity violation of weak interactions in atomic nuclei due to anapole concept produced by static currents. Dynamic toroidal moment is less known. Although it radiates electromagnetic fields with angular momentum equal to conventional dynamic multipole momenta, toroidal moments are often excluded from the standard multipole expansion of current excitation [3,4].

Interestingly, omitting the toroidal moment in the multipole expansions leads to nonphysical results in media with toroidal topology [5,6], although the toroidal moment in multipole expansion does not usually contribute significantly to radiation stemming from sources with dominating electric and magnetic moments.

Experimentally, toroidal moment was detected quite recently in metamaterials. These artificial materials exhibit electromagnetic properties which do not occur in natural materials. In particular, they present a unique opportunity for manipulating features on the subwavelength scale, allowing thus to achieve effects such as negative refraction, cloaking, strong field localization, etc [7-11]. Kaelberer et al. were able to induce currents in Split Ring Resonators (SRR) arrangements, resembling poloidal currents flowing along the meridians of the torus due to the configuration of conductive currents in the SRR [12]. This configuration allowed for strengthening of the toroidal moment to detectable level and, simultaneously, weakening of the electric and magnetic moments. Observation of the toroidal response in specially designed three-dimensional clusters caused a significant impact on the metamaterial community [2,12-18] and gave rise to a series of exciting discoveries in the field of toroidal electrodynamics (see Ref. 2 and references therein).

A distinct feature of toroidal metamaterials, the anapole excitation, is of non-trivial, non-radiating nature. In particular, the term «non-trivial», refers here to the ability of radiating the vector potential in the absence of electromagnetic fields. Moreover, at the same time, several publications emphasize the properties of destructive interference between the toroidal and the electric dipole moments. The result of this interference is the reduction of the radiation losses in metamaterials that produces an effect analog to electromagnetically induced transparency (EIT) [14-16]. Several works discussed the possibility of radiating vector potentials in the absence of electromagnetic fields leading to the dynamic Aharonov-Bohm effect [3,14,5,17]. These works claimed a high Q-factor associated with toroidal metamaterials to be due to the anapole excitations. This is significant for cloaking behavior and many other applications in photonics and plasmonics demanding strong localized fields, such as nonlinear excitations, high Q-factor cavities of spacers, lasers, qubits [2,15,17,18,19]. As mentioned above, anapole sources might also offer important applications for the dynamic Aharonov-Bohm effect. This problem is of interest for many reasons; one of them is the prospect for secure data communication [3,5,14, 17,18].

We note, that the toroidal topology of metamolecules (the elementary blocks of this kind of metamaterials) is complicated because of the necessity to design elements resembling the toroid geometry [2]. This requirement limits the application of toroidal metamaterials in the visible and THz frequency range, where the toroidal metamolecules could be fabricated as 3D inclusions at the micro- and nano-scale. The planar metamolecules can be fabricated easier, e.g. by photolithography [20].

Traditional planar metamaterials based on SRRs and their hybrid modifications have been extensively studied in recent years for demonstrating negative refraction, magneto-inductive waves, THz modulators and biological/chemical sensors [21-28].

Exotic properties of hybrid SRRs are observed in connection with the resonant nature of their response. Both the electric and magnetic responses are accompanied by strongly localized electromagnetic fields within the metamolecules. However, the Q-factor of such metamaterials is limited by radiation losses associated with the fields scattered by metamolecules.

There are several approaches to minimize the radiation losses in metamaterials. The first method is to use asymmetric SRR with the possibility of exciting two destructively interfering bright modes. In particular, Fedotov et al. proposed two asymmetric SRRs with slightly different shapes. The radiation losses are rather low here due to the dark mode caused by destructive interference between currents in each SRR [29]. This interference leds to an asymmetric peak in transmission characteristics and a high Q-factor. At the same time, such SRR elements need to be almost overlapping, which implies strict requirements for fabrication.

The second method is known as the EIT (electromagnetically induced transparency) in metamaterials and involves the excitation of two modes, bright and dark. This approach allows to couple a radiative bright mode with a subradiative and non-interacting with plane wave dark mode. In this case the hybrid metamolecule consists of two elements, each one of them supporting these modes. Hybridization of the resulted resonances produces a narrow peak of the Fano-type resonance which was considered previously in atomic systems [30, 31]. The concept of metamaterials allows for observing EIT in many artificial structures [32-34].

The third approach, based on the anapole excitation, is known as an analog of EIT in toroidal metamaterials [2,14,5]. In contrast to the Fano-type resonance, the anapole excitation does not require two scattering channels in metamolecules. It is generally accepted that anapole is a mode that occurs as a result of destructive interference between the toroidal and electric dipole moments, which are both radiating [3,6,14,17]. For the ideal anapole, radiation losses are absent due to the above mentioned interference in the far-field zone. Thus, one can expect a high Q-factor in anapole metamaterials.

In this paper, we theoretically and experimentally study planar anapole metamaterials. We show that the anapole metamolecule is an ideal resonator with an extremely high Q-factor accompanied by strong localization of the electromagnetic fields within the metamolecule.

## II. THE POINT ANAPOLE RESONATOR

Let us consider first the toroidal dipole $T$ source given by the following formula in terms of the current density $j$: (We shall consider the limit as the source approaches the point-like limit [3], and a harmonic excitation of the form $exp(i\omega t)$)

$$\boldsymbol{T} = \frac{1}{10c} \int d^3 r (\boldsymbol{r}(\boldsymbol{r} \cdot \boldsymbol{j}) - 2\boldsymbol{j}r^2) \quad (1).$$

At the same time the electric dipole is given by the well known formula:

$$P = \frac{1}{i\omega} \int d^3 r \, j \quad (2).$$

A distinctive feature of the toroidal dipole is its ability to radiate with the same angular momentum as the electric dipole [3]. Indeed, consider the toroidal and electric dipoles placed at the origin, $r=0$. One can see that the electric and magnetic fields radiated by superposition of the two dipoles are:

$$E_{tot} = E_P + E_T = \left[\frac{r \cdot (P - ikT)F(\omega,r)}{c^2 r^2} r - \frac{G(\omega,r)}{c^2}(P - ikT)\right]\frac{\exp(-ikr + i\omega t)}{r} \quad (3a)$$

$$H_{tot} = H_P + H_T = -\frac{ikD(\omega,r)}{cr}[r \times (P - ikT)]\frac{\exp(-ikr + i\omega t)}{r} \quad (3b),$$

where $D, F, G$ can be found in [3,14].

Note, that the fields of anapole disappear in the case of $P=ikT$, i.e. destructive interference of toroidal and electric dipole moments takes place everywhere except $r=0$ [3,14]. This configuration forms non-radiating point anapole, with the fields existing only at the point $r = 0$, and described by the $\delta$ – function [3,17]:

$$E_{tot}(r = 0) = ikT\delta(r)\exp(i\omega t) \quad (4a)$$

$$H_{tot}(r = 0) = ik \, \text{rot}(T)\delta(r)\exp(i\omega t) \quad (4b).$$

It makes sense to describe the topology of point anapole [17]. Equations (4a,b) determine a localized electric field at the point of origin ($r=0$) while the magnetic field is represented by a loop in the orthogonal to $E$ plane. Without dissipation losses, this allow us to design a resonator, which is free of radiation losses; the entire power is confined at only one point, which corresponds to the point-like distribution of currents, proportional to $\delta(r)$. It also means that the anapole is an ideal resonator with an infinite Q-factor. A high Q-factor resonator has a minimum loss power $P_d$, i.e. $Q=\omega_0 W/P_d$, where W is the energy stored in the resonator. Obviously, our intention will be to apply the resulting property for the realistic case of a practical configuration. Finite size current density can be expanded in powers of $a/r$, where $a$ is the characteristic size of the source. In practice, other parasitic multipoles besides toroidal and electric dipole moments will be excited and will contribute to a significant reduction of the Q-factor as a result of their radiation. To reduce the role of the parasitic multipoles and boost the anapole contribution, the resonator design needs to maintain a spatially confined magnetization, circulating around a concentrated electric field. It is important that the volume occupied by electric and magnetic fields tends to zero within the metamolecule. The field distributions, different from the $\delta$-function, will be accompanied by the contribution of additional dipole moments thus further reducing the Q- factor.

Metamaterials that support a strong toroidal response are well known. Usually they consist of rather complicated 3D metamolecules, although some of them are based on SRR modified inclusions [2]. Fabrication of 3D metamolecules is challenging, especially in the visible and THz spectrum. In this regard, the need for a design of a planar toroidal metamaterial is evident, especially for the anapole metamaterials.

### III. THE STRUCTURE OF THE SYSTEM

Here we report metamaterials consisting of planar conductive metamolecules. Each metamolecule is formed by the two symmetrical split rings (inset on Fig. 1). The incident plane wave with electric field **E** paralled to the central wire excites circular currents **j** along the loops. Each current induces the circulating magnetic moments **m** wreathing around the central part of metamolecule. As a result, this leads to a toroidal moment **T** oscillating back and forth along the axis of the metamolecule. Two side gaps also support a magnetic quadrupole moment **Qm**. Moreover, due to the central gap electric moment **P** can be excited in the metamolecule (Fig. 1);

the central gap is a necessary part of anapole. We aim to miniaturize metamolecules in order to bring their field close to the geometry of a point anapole.

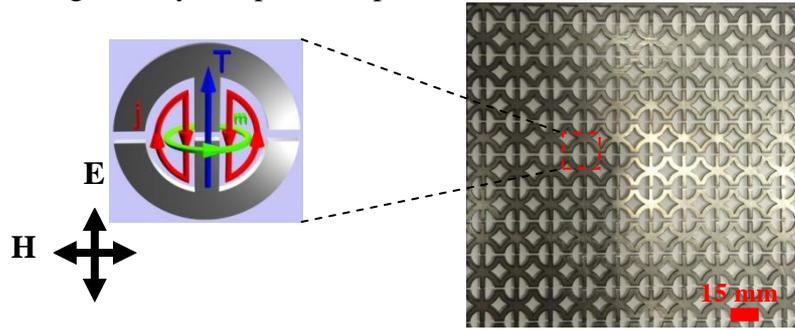

Fig. 1 A fragment of a metamaterial supporting toroidal dipolar excitation. Red arrows show displacement currents *j* induced by the vertically polarized plane wave, blue arrow shows toroidal dipole moments **T** of the metamolecule, green arrow shows circulated magnetic moment **m**; the picture shows a metamaterial sample.

We performed simulation of the structure depicted on Fig.1 by the commercial solver CST Microwave Studio. Simulated structure consisted of metamolecules placed at a distance 15 mm from each other. The internal gap is 0.75 mm wide, the outer gaps are 1 mm, and the metal thickness is 2 mm. The extremely narrow resonance appearing as a dip in the calculated transmission curve (S21- parameter) is observed at around 9.54 GHz, see Fig. 2. We note that the achieved Q-factor is very high ~ $3.817 \times 10^6$, that corresponds to 2.5 KHz linewidth at 3dB level above the bottom of the dip.

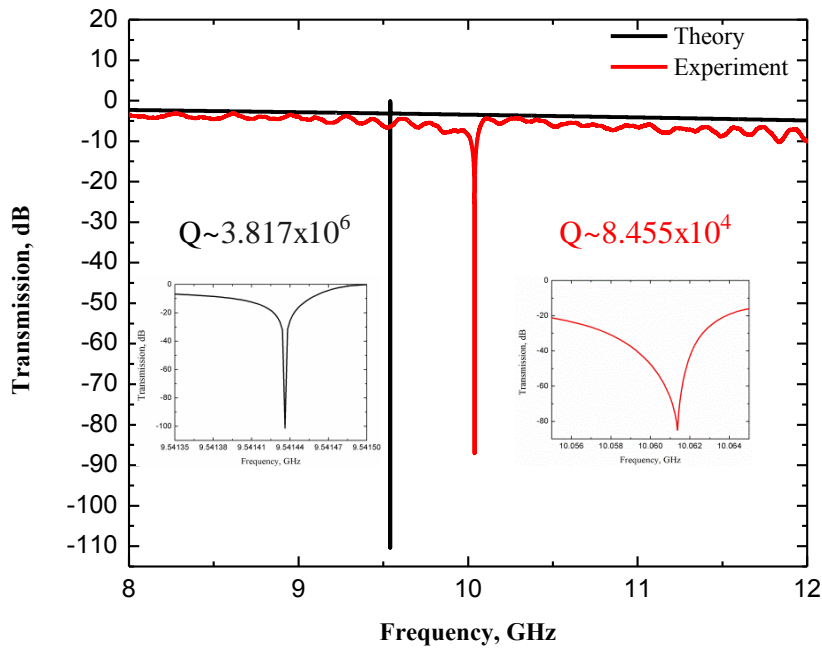

Fig 2. Theoretical results of transmission calculated by CST Microwave Studio (black lines) and experimental (red lines) spectra obtained for the sample shown in Fig. 1.

The anapole character of this extremely narrow resonance is confirmed by calculating the distribution of the local fields and the density of the displacement currents induced in the metamolecule, which are depicted in Fig. 3. The magnetic field at the resonance frequency corresponds to the closed vortex circulating around the central axis of the metamolecule. At the same time, the electric field is localized in the central gap in a region of about $(1/50)\lambda$ and is related to the field $E_0$ of the incoming plane wave as $E/E_0 = 19160$, whereas magnetic field is related to $H_0$ as $H/H_0=688$. We note that this configuration of fields is close to the topology of point anapole proportional to a $\delta$- function.

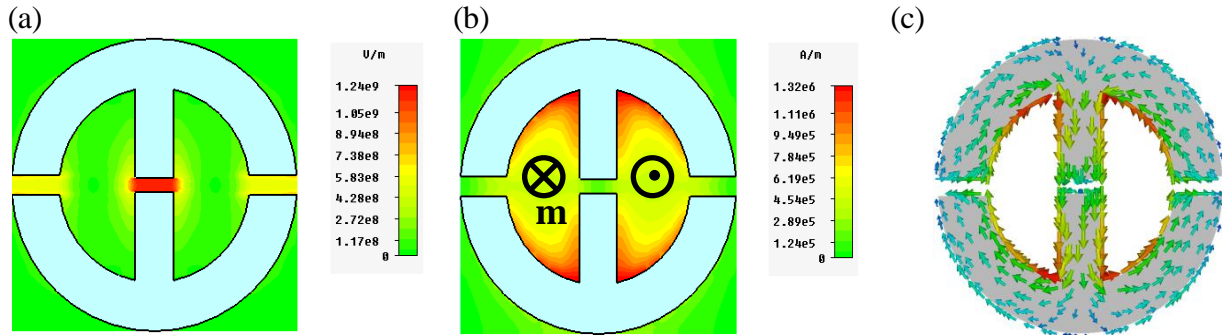

Fig. 3 Calculated distributions of the corresponding electric field |*E*| (a*)*, magnetic field (absolute value |H| (b)), and amplitude of the conductive current **j** (c) induced in the metamolecule at 9.54143 GHz.

To access the role of the anapole contribution in forming the observed response, we calculate the relative strength of the standard multipoles in terms of the electromagnetic power they scatter in the far-field zone. The multipole moments induced in the metamolecules are calculated based on the density of the conducting currents in metamolecules (Fig. 4). This approach allows us to clearly reveal the near field signature of the multipolar current excitations to their electromagnetic response in the far field. One can see that at the resonance, the contribution of the magnetic dipole and electric quadrupole are strongly suppressed and there is a narrow range of frequencies close to 9.55 GHz, where the far-field scattering due to the resonant toroidal and electric excitations dominate all other standard multipoles. It is important that in the vicinity of the frequency f =9.54143 GHz the power radiated by the toroidal dipole **T** prevails in the system and is equal to the power of the electric dipole moment **P**. This corresponds to the anapole excitation in metamolecules and confirms the field configuration shown in Fig. 3. At the same time, **T** and **P** are more than $10^3$ times greater than the other multipoles in the system with the exemption of the magnetic quadrupole **Qm** which is quite high. Interestingly, it was concluded in [20] that isolated toroidal dipole moment is in principle unachievable in planar geometry and strong magnetic quadrupole moment **Qm,** as well as electric octopule moment **Oe** always accompanied metamaterial response. The contribution of the magnetic quadrupole **Qm** is fully distinguished from the anapole contribution in the metamolecule. Its manifestation is obvious and completely corresponds to the Savinov's consideration [20]. The presence of quadrupole **Qm** is a parasitic factor and prevents us to achieve the even higher Q-factor predicted by the point anapole concept. Nevertheless, to our knowledge, the numerically computed Q = $3.817 \times 10^6$ is the highest value ever achieved by modelling of a metamaterial.

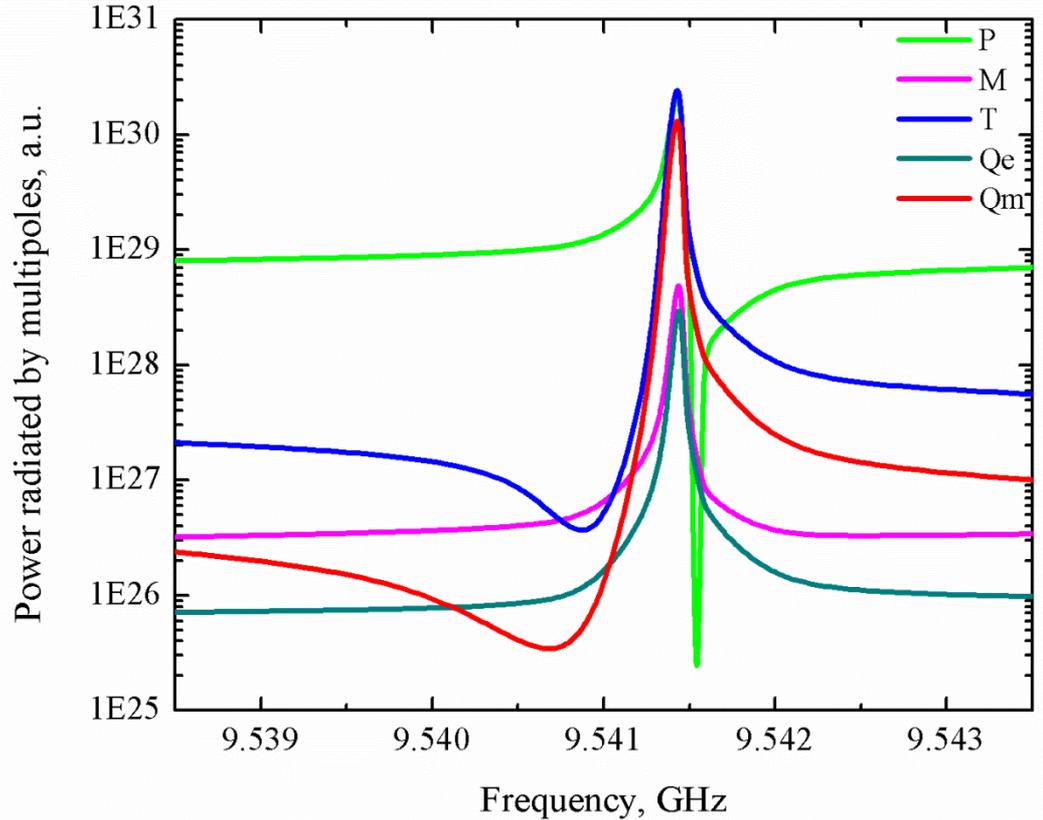

Fig. 4. Contributions of the five strongest multipolar excitations (see text) to the reflection of the metamaterial array. It is obtained by coherent (amplitude and phase) summation of all multipole contributions. The log scale in the y axis is chosen so as to reveal more clearly the contribution of the quadrupole terms as well.

Let us mention now other types of resonators that achieve high values of Q-factor and are classified as Ultrahigh Q resonators according to the standards in optical resonators. These resonators possess complicated bulk geometry and have dimensions higher than that corresponding to the resonance wavelength. They are based on principles such as the Fabry-Perot resonances (Q~2000), the Photonic crystals (~15000) and the whispering gallery ideas (Q~$10^9$, although with the volume of cavity ~3000 $\mu m^3$) [35 and references therein]. The highest Q-factor planar resonator (Q~$10^6$) known as superconductor spiral resonator has effective dimension more than a wavelength [36].

We note that planar metamaterials have important limitations as dissipation losses of metamolecules and substrate (in particular, due to Fabry- Perot resonances at certain frequencies such that the wavelength in the dielectric fits in the substrate thickness); as a result, multiple reflection of electromagnetic waves occurs between the walls of the substrate [20].

To demonstrate the possibilities of the proposed metamaterial, we have fabricated the sample by laser cutting method from a slab of steel. We have skipped the dielectric substrate to avoid the loss factors associated with it. We included in the sample 10x10 metamolecules of 2 mm thickness; each metamolecule is cut in the form depicted on Fig. 1. The metamolecules are located at a distance of 15 mm from each other.

### IV. EXPERIMENTAL CHARACTERIZATION OF METAMATERIAL SAMPLE

For experimental characterization of metamaterials we carried out measurements of *S21*-parameters (transmission) in an anechoic chamber by two horns method. Two broadband horn antennas P6-23M for electromagnetic radiation emission and detection were located at a distance 1 m from the metamaterials samples. The transmission coefficient *S21* of the electromagnetic waves through the metamaterial slab was measured by a vector network analyzer Rohde & Schwarz SVB20 at frequencies 8-12 GHz.

We would like to stress here relevant experiment details. A narrow resonance peak or dip is expected to be sensitive to a variety of external factors: vibrations, multiple reflections, and, especially, the position of the metamaterial sample relative to the distribution of the external field. For this reason we have applied micrometer screws for accurate adjustment of the sample position. Thus, we found a high Q-factor resonance dip close to the frequency of 10.06 GHz (Fig. 2, red line). To identify this narrow deep, we used the 40,000 points in a selected frequency range. A sharp resonance dip (at half-power above the minimum transmission its width was 119 KHz) was found at the central frequency of 10.061 GHz. This value is rather close to that of 9.54143 GHz expected from simulations. The inset in Fig.2 shows an expanded graph of the dip in the frequency range 10.0560-10.0650 GHz. Thus we can estimate the Q-factor of the resonance as $8.455 \times 10^4$. The origin of deep is defined as the anapole mode (**T**+**P**) accompanied by a parasitic magnetic quadrupole **Qm**.

Next we discuss possible applications of anapole metamaterials. Anapole high Q-factor metamaterial proposed in this paper offers interesting opportunities in connection with the light-matter interaction. Tunable metamaterials and modulators at the THz range are accompanied by interaction of strong fields localized at metamolecules and the surfaces of semiconductor inclusions; the latter are located near metamolecules and can act as tunable elements [37]. The tunability appears here because the conductivity of semiconductors can be varied by an external femtosecond pump laser or by electrical gating. The semiconductor inclusion can be metamaterial substrate or an additional inclusion in gap of the SRR where the ac electric field has its maximum amplitude. However, THz modulators and sensors achieving strong tunability and sensitivity are limited usually by the low Q-factor of metamaterials, mainly due to radiation losses associated with the electric and magnetic dipolar responses of metamolecules. Thus, metamaterials with dominating anapole response should be capable of supporting extremely high Q-factors and are perfect candidates for modulators achieving strong tunability with smaller pump power. Since the region of localized electric field is of subwavelength extent, one can reduce device dimensions for so-called THz spectral regime, for which less pump power is required for tuning the properties of semiconductors.

Another possibility for anapole response application is in reducing radiation loss of superconducting quantum bits [39,40] and, in particular, in qubits playing the role of meta-atoms in quantum metamaterials [19]. When cooled to ultra-low temperatures, superconducting loops containing Josephson junctions work as macroscopic two-level quantum systems, commonly called as qubits. Reducing radiation loss of qubits makes their coherence times higher, which brings an advantage for qubit functionality.

We note that quantum nonlinear resonators, in order to act as qubits, need to have as low as possible radiation losses. Therefore, in order to increase the lifetime of the qubit, resonators should have a high Q-factor and should be miniaturized. Increasing the size of meta-atoms to the dimensions comparable to the wavelength leads to radiation losses. Anapole meta-atoms are very promising candidates for being used as qubits, providing significant benefits in this area due to their high Q-factor and reduced radiation losses.

In summary, we have designed and fabricated an anapole metamaterial consisting of planar metamolecules. The anapole behavior stems from the fact that the multipole radiation terms interfere destructively. As a result, a very high Q-factor obtained at the resonance and is limited mainly by higher-order effects. Such an anapole behavior and the corresponding high Q-factor can find their use in modulators and sensors as well as in superconducting qubits, substantially increasing their quantum coherence times.

**ACKNOWLEDGEMENTS**

The authors gratefully acknowledge the financial support of the Ministry of Education and Science of the Russian Federation in the framework of Increase Competitiveness Program of NUST «MISiS» (№ К4-2015-031), the Russian Foundation for Basic Research (Grant Agreements No. 16-32-50139 and No. 16-02-00789), the European Union in the framework of the project GA -320081 PHOTOMETA, as well as the Deutsche Forschungsgemeinschaft (DFG).